\documentclass[12pt]{article}
 \usepackage[dvips]{graphicx}
\begin{document}
\title {On Some Physical Aspects of Planck-Scale Relativity:A Simplified Approach}
\bigskip
\author{~ Alex Granik\thanks{Department of Physics, University of the Pacific,
Stockton,CA.95211;~E-mail:agranik@uop.edu}}
\date{}
\maketitle
\begin{abstract}
The kinematics of the two-scale relativity theory (new
relativity) is revisited using a simplified approach. This
approach allows us not only to derive the dispersion equation
introduced earlier by Kowalski-Glikman, but to find an additional
dispersion relation, and, even more important, to provide a
physical basis for such relations. They are explained by the fact
that in the observer invariant two-scale relativity ( new
relativity) the Planck constant does nor remain constant anymore,
but depends on the universal length scale. This leads to the
correct relation between energy and frequency at any scale.

\end{abstract}
\section{Introduction}
The pioneering papers By G.Amelino-Camelia \cite{GA1},\cite{GA2}
have explicitly and unambiguously introduced what can be called a
"two-scale special relativity" (or a new relativity).
 The crucial feature of this new relativity is the introduction of an
observer-independent absolute scale (possibly the Planck scale
$\Lambda \sim 1.6 \times 10^{-35}m)$ in addition to the existing
in special relativity  observer- independent universal velocity
scale, the speed of light. The investigation of  a structure of
the emerging space-time at the Planck scale ( different from
Minkowski space-time) is based upon $\kappa$-Poincare algebra
(\cite{JL} and references therein). It should be mentioned that
consequences of the finite universal length scale on the
resulting mechanics have also been studied by L.Nottale,
C.Castro, and others researchers (e.g., \cite{LN}, \cite{CC},  and
references therein),albeit from a different point of view. \\

J.Kowalski-Glikman \cite{JG} investigated the space-time
transformations and found that the metric structure of the
respective space-time events has the Minkowski distance as an
invariant. He also proposed the dispersion relation which removes
a seemingly paradoxical discrepancy between the infinite group
velocity of a wave and the finite velocity of a massless
particle. However the author emphasized that this relation is
required a justification "by more solid arguments".\\

We have to emphasize that here we are not going to dwell on the
algebra ($\kappa$-Poincare) underlying the respective space, but
rather try via a simplify approach to revisit some physical
points of the problem.\\

In view of this, we provide a simplified derivation of the
dispersion relation and show that there is an $additional$
dispersion relation, consistent with the finite velocity of  a
massless particle. We also provide an explanation of the physical
reason behind these relations.

The first of these relations corresponds to the boost
transformation of position and time carried out in the space of
the boost parameter $\xi$, independent of the universal invariant
length scale $\lambda$. In this case the transformation does not
coincide with the special relativistic transformation, and
neither $p_0$ (energy) is proportional to the frequency nor the
momentum $p_{\alpha},~\alpha =1,2,3.$ is proportional to the wave
number.

On the other hand, the second transformation carried out in the
space of the boost parameter $z$ which is  a function of the
length-scale-independent boost parameter $\xi$ and the invariant
length scale itself. In this space the transformation coincides
with the special-relativistic boost,  the momentum is
proportional to the wave number, but energy is not proportional to
the frequency, requiring a certain modification of the relation
between them.

The explanation of this situation ( and equally of the situation
with the previous case) can be as follows. Since in quantum
mechanics energy is given by $\hbar\times frequency,$ and now
frequency is not proportional to energy, this means that the
introduction of the additional invariant parameter ( invariant
length scale $\lambda$) into a description of mechanics results in
other parameters ( including $\hbar$) becoming functions of
$lambda$. This leads to the emergence of the so-called
$effective$ Planck constant(e.g. ref\cite{CC}) and the
accompanying restoration of the proportionality between energy
and frequency.

In addition, by examining the metric structure of the space-time
with the help of  simple arguments we explicitly find the
boundaries of time-like/space-like intervals. It has turned out
that these boundaries indicate two mutually exclusive
possibilities:
\begin{center}
            1)$physical$. Massive particles cannot reach the speed of light,
            in a world with two independent universal observer-independent
            scales,\\

            2)$hypothetical.$ Massive particles can move with velocities reaching
            the speed of light, but two universal scales of the theory
            are not independent.
\end{center}
Finally, we investigate the uncertainty relation connected with
the 2-scale relativity and compare it with the respective
relation in the string theory.\\

\section{Dispersion Relation in Planck-Scale Relativity}
We begin our analysis by considering the Minkowski distance
(written in the differential form) , which we assume to be
invariant as in special relativity:
\begin{equation}
\label{eq:1} \frac {dx_0^2}{ds^2}-\frac{dx_{\alpha}^2}{ds^2} =
\frac{dx_0'^2}{ds^2}-\frac{dx_{\alpha}'^2}{ds^2}
\end{equation}
where $x_0,~ x_{\alpha}, ~\alpha=1,2,3$ are the time and space
components of the four-vector $x_i (~ i=0,1,2,3)$, the primes
denote a transformed (under the boost) system, and $ds$ is the
appropriate space-time interval.

In the system at rest $dx_{\alpha}^2/{ds^2}=p_{\alpha}=0$ and
\cite{JG}
\begin{equation}
\label{eq:2} (\frac{dx_0}{ds})^2 =
[\frac{1}{\lambda}sinh(p_0\lambda)]^2
\end{equation}
Here $p_0$ and ${p_{\alpha}}$ are the energy and momentum
components of the momentum four-vector $p_0,p_\alpha$. This means
that in the rest frame the interval $ds$ has the following
dependence on $dx_0$:
$$ds=dx_0\frac{\lambda}{\sinh(m\lambda)}$$where $m$ is the mass
(rest energy). From  the  relation between $p_0$, $p$, and $m$
found in \cite{NB} on the basis of the group-theoretical arguments
\begin{equation}
\label{eq:3}
[\frac{2}{\lambda}sinh(\frac{m\lambda}{2})]^2=[\frac{2}{\lambda}
sinh(\frac{p_0\lambda}{2})]^2 -p_{\alpha }^2e^{p_0\lambda}
\end{equation}
we find that in the rest frame
\begin{equation}
\label{eq:4}
[\frac{1}{\lambda}sinh(\frac{p_0\lambda}{2})]^2=[\frac{2}{\lambda}sinh(\frac{m\lambda}{2})]^2
\end{equation}

Using (\ref{eq:4}) in (\ref{eq:1}) we obtain the following
expression for the time-space interval $ds$.
\begin{equation}
\label{eq:5} (ds)^2
=[\frac{\lambda}{sinh({m\lambda})}]^2(dx_0^2-dx_{\alpha}^2)
\end{equation}

For the following we drop the prime denoting the transformed
system and introduce action $S$ for a free particle similar to
the one in special relativity:
\begin{equation}
\label{eq:6} S=-a\int ds = -a\frac{\lambda}{sinh({m\lambda})}\int
dx_0\sqrt{1-V_{\alpha}^2}
\end{equation}
where $a$ is some constant to be determined later.

The Lagrange function for the particle is then
\begin{equation}
\label{eq:7} {\cal L}=-a\frac{\lambda}{sinh({m\lambda})}
\sqrt{1-V_{\alpha}^2}
\end{equation}
Here $V_{\alpha}$ is a component of the three-velocity of the
particle ( in the particular frame).\\

Using (\ref{eq:7}) we find the momentum $p_{\alpha}$ of the
particle
\begin{equation}
\label{eq:8} p_{\alpha}= \frac{\partial {\cal L}}{\partial
V_{\alpha}}=a\frac{\lambda}{sinh({m\lambda})}\frac{V_{\alpha}}{\sqrt{1-V_{\alpha}^2}}
\end{equation}
If we introduce a reduced mass
$$\bar{m}=\frac{sinh(m\lambda)}{\lambda}$$ then Eq.(\ref{eq:8})
looks exactly as the respective relation in the conventional
special relativity:
\begin{equation}
\label{eq:9} p_{\alpha}=\frac{\bar
{m}V_{\alpha}}{{\sqrt{1-V_{\alpha}^2}}}
\end{equation}

{\bf I}. First, we consider all the relations expressed in terms
of the length-scale independent boost parameter $\xi.$ With this
in mind, let us boost a massive particle (of mass $m$), initially
at rest. Its  momentum $p_{\alpha}$ \cite{NB} is then
\begin{equation}
\label{eq:10} p_{\alpha}=\bar{m}~ sinh\xi\frac{\hat{p}_{\alpha}}
{\lambda\bar{m}~ cosh\xi+\sqrt{1+\bar{m}^2\lambda^2}}
\end{equation}
where $\hat{p_{\alpha}}$ is the unit vector of $p_{\alpha}$.\\

Comparing (\ref{eq:8}) and (\ref{eq:10})we obtain
\begin{equation}
\label{eq:11}
a\frac{V_{\alpha}}{\sqrt{1-V_{\alpha}^2}}=\hat{p_{\alpha}}
\bar{m}^2\frac{sinh\xi}{\lambda \bar{
m}cosh\xi+\sqrt{1+(\bar{m}\lambda)^2}}
\end{equation}
From the dimensional considerations we set the value of $a$
$$a=\bar{m}^2.$$ Solving  (\ref{eq:11}) with respect to $V_{\alpha}$ we
get the following two cases depending on the sign of $\lambda$:

$a)$ $\lambda >0$
\begin{equation}
\label{eq:12} |V_{\alpha}|= \frac{sinh\xi}{cosh\xi
\sqrt{1+\lambda^2\bar{m}^2}+\sqrt{1+\bar{m}^2\lambda^2}}
\end{equation}\\
This is exactly the expression obtained in \cite{JG} on the basis
of different considerations. For the future analysis we denote
the upper bound on the velocity $|V_{\alpha}|,$  reached in this
case at $\xi\rightarrow \infty$,  as the speed of  "light"
$c_{\lambda}:$ $$c_{\lambda}\equiv 1/cosh(\lambda m)\equiv
1/\sqrt{1+\bar{m}^2\lambda^2}$$

From (\ref{eq:12}) we can easily see that time-like ( and
space-like) regions of the respective Minkowski space for a
massive particle are not fixed anymore, but depend on its
mass:$$x_0=x\sqrt{1+(\bar{m}\lambda)^2}.$$ Amusingly enough, with
the increase of mass $m$ the time-like region decreases, and in
the $hypothetical$ case of $m\lambda \rightarrow \infty$  shrinks to zero!\\

$b)$ Another case corresponds to the values $\lambda \leq 0$. In
this case the limiting value of $|V_{\alpha}|$ can be equal to
$1$: $$|V_{\alpha}|=\frac{sinh\xi}{cosh\xi
\sqrt{1+(\bar{m}\lambda)^2}-\bar{m}|\lambda|}=1$$

From this expression follows that the limiting value of $\xi$ is
$$cosh(\xi)=coth(m|\lambda|)\equiv
\frac{\sqrt{1+\bar{m}^2\lambda^2}}{\bar{m}|\lambda|}\equiv\frac{1}
{\sqrt{1-c_{\lambda}^2}}.$$ The respective value of the momentum
$p \rightarrow \infty$ as in the case of the conventional special
relativity. What is most interesting about this case that now two
scales, speed of light and $\lambda$, are not independent,
hinting at the existence of only one observer-invariant scale,
length scale $\lambda.$\\

Since expression (\ref{eq:9}) for $p_{\alpha}$ as a function of
$V_{\alpha}$ formally looks exactly like its special relativistic
counterpart, this gives us a clue for undertaking the task of
casting the appropriate equations into the form found in special
relativity. We begin with the relation between energy $p_0$
$^{\cite{NB}}$ :
\begin{equation}
\label{eq:13} p_0=\frac{1}{\lambda}log[~\bar{m}\lambda~cosh\xi
+\sqrt{1+(\bar{m}\lambda)^2}~]
\end{equation}
and the momentum $p_{\alpha}$ given by Eq.(\ref{eq:10}).\\

Since the identity $cosh^2\xi -sinh^2\xi\equiv 1$ (yielding in
particular the relation $E^2=p^2+m^2$) allows to reach our goal,
we have to express $cosh$ and $sinh$ in terms of $p_0$ and
$p_{\alpha}$ using (\ref{eq:10}) and (\ref{eq:13}). For
convenience sake, in the following I will drop the subscript at
$p_{\alpha}.$ Comparing (\ref{eq:10}) and (\ref{eq:13}), we find
that
\begin{equation}
\label{eq:14} sinh\xi=\frac{pe^{p_0\lambda}}{\bar{m}}
\end{equation}
On the other hand, from (\ref{eq:13}) we obtain
\begin{equation}
\label{eq:15} cosh\xi=\frac{1}{\bar{m}}\frac{e^{p_0\lambda}-
\sqrt{1+\bar{m}^2\lambda^2}}{\lambda}
\end{equation}\\

Now it is easy to identify what we call  the {\it effective
frequency} $\bar{\omega}$ and the $effective~~ wave ~number ~~
\bar{k}$:
\begin{eqnarray}
\label{eq:16} \bar{k}=pe^{p_0\lambda}\\\nonumber\\
\bar{\omega}=\frac{e^{p_0\lambda}-
\sqrt{1+\bar{m}^2\lambda^2}}{\lambda}
\end{eqnarray}\\

According to (\ref{eq:14}) and (\ref{eq:15}), the  relation
between them is the exact copy of the respective relation in
special relativity:
\begin{equation}
\label{eq:18} \bar{\omega}^2=\bar{k}^2+\bar{m}^2
\end{equation}\\
Thus we have arrived at the dispersion relation by identifying the
frequency  not with the energy part $p_0$ of the four-momentum
but with a certain function (to be found) of energy and momentum,
which we denote by $\bar{\omega}$, and using the same procedure
for the momentum part $p$ ( by identifying the wave number
with a certain function $\bar{k(p,\lambda)}).$\\

{\it{The physical justification of this is as follows. On the
scales much exceeding $\lambda$ the Planck constant is not
affected by the existence of the universal length scale. However,
at the scales comparable with $\lambda$ this is not true anymore,
and as a result the Planck $"constant"$ becomes the $effective$
Planck function $\hbar_{eff}$ which varies with $\lambda$,
momentum $p$, and energy $p_0$ (e.g., reference  \cite{CC}). In
the limit of $\lambda\rightarrow 0~~~ \hbar_{eff} \rightarrow
\hbar.$ Thus despite the fact that now frequency is a function of
$p,~p_0, ~ and~ \lambda~$ the overall linear relation between
energy $p_0$ and frequency is preserved. This means that  since
$\omega =p_0$ (in the units where the Planck constant $\hbar =1$):
\begin{equation}
\label{eq:19} energy=h_{eff}^0p_0=\bar{\omega};~~
momentum=h_{eff}^1p=\bar{k}
\end{equation}\\

What seems very interesting is that the emerging effective Planck
numbers ( or more correctly functions) in such 2-scale relativity
have to transform differently for the spatial and temporal parts
of the four-momentum to be consistent with the fact that
$V_{m=0}=1$. It looks as if the Planck function also becomes sort
of 4-vector. We investigate the relations (\ref{eq:19}) in more
detail in the next section.}}

Now the group velocity $v_g$ $$v_g=\frac{d\bar{\omega}}{d\bar{k}}
= \frac{\bar{k}}{\sqrt{\bar{m}^2+\bar{k}^2}}$$is $always$ less,
and the phase velocity $v_p$
$$v_p=\frac{\bar{\omega}}{\bar{k}}=\frac{\sqrt{\bar{m}^2+\bar{k}^2}}{k}$$
is $always$ greater than the maximum speed $c_{\lambda}$
attainable by a massive particle
$$c_{\lambda}\equiv\frac{1}{cosh(m\lambda)}=
\frac{1}{\sqrt{1+(\bar{m}\lambda)^2}}.$$ For a massless particle
both the group velocity and phase velocity reach the speed of
light, in full agreement with the expression for $V_{\alpha}$
(\ref{eq:12}). The well-known theorem of the conventional
relativity stating that
$$v_{group}^2v_{phase}^2=1$$ holds true. \\

The effective frequency $\bar{\omega}$ is given in terms of $p_0$
and $\lambda m$, Eq.(17). Using the basic relation (\ref{eq:3})
between $m$, $p$, and $p_0$ we derive the relation between
$\bar{\omega}$ , Eq.(17) and $p_0$ by inserting the value of
$cosh(m\lambda)\equiv \sqrt{1+(\bar{m}\lambda)^2}~$ from
(\ref{eq:3}) into (17). This would immediately give us the
following
 \begin{equation}
 \label{eq:20}
 \bar{\omega}= \frac{sinh(\lambda p_0)}{\lambda}+
 \frac{\lambda p^2e^{\lambda p_0}}{2}
 \end{equation}\\

The derived relations (\ref{eq:16}) and (\ref{eq:20}) are the
relations introduced in \cite{JG} without a proof. Moreover, we
have provided a physical justification for these relations.\\

Now we will recast the expressions for the spatial part of 4-
velocity (\ref{eq:12}) into the form similar to special
relativity. Recalling that in special relativity $sinh\xi$
corresponds to the spatial part $u$ of the four-velocity,
equation (\ref{eq:14}) can be written in terms of the $reduced$
four-velocity $u$:
\begin{equation}
\label{eq:21} \frac{\bar{p}}{\bar{m}}=sinh\xi=\bar{u}
\end{equation}
We  express $\bar{u}$ in terms of $V$ with the help of
(\ref{eq:12}), use the definition of the "speed of light"
$c_{\lambda}$, and denote $pe^{\lambda p_0}=\bar{p}.$ As a
result, we get
\begin{equation}
\label{eq:22}
\bar{u}=\pm\frac{Vc_{\lambda}}{\sqrt{1-V^2}\mp\sqrt{1-c_{\lambda}^2}}
\end{equation}
In the limit of $c_{\lambda}\rightarrow 1$ (that is in the limit
yielding the conventional special relativity) we get the familiar
expression for the spatial part of the four velocity in special
relativity: $$\bar{u}=u =\pm\frac{V}{\sqrt{1-V^2}}.$$ Note the
symmetric character of the "forward" ($+$) and "backward" ($-$)
velocity in this case. This feature is lost in the general
expression
(\ref{eq:22}).\\

In another limit of $V\rightarrow c_{\lambda}$ we get
\begin{eqnarray}
\label{eq:23} \bar{u}^+ \rightarrow \infty \nonumber
\\
\bar{u}^-\rightarrow
-\frac{c_{\lambda}^2}{2\sqrt{1-c_{\lambda}^2}}
\end{eqnarray}
indicating that the region of "backward" velocities is bounded
from below for all the values of $c_{\lambda}\neq 1.$ \\

Using the definition of the reduced velocity according to
$$\bar{V}=tanh\xi=\frac{\bar{u}}{\sqrt{1+\bar{u}^2}}$$ we easily
obtain expression for $\bar{V}$ as a function of $V$:
\begin{eqnarray}
\label{eq:24}
\bar{V}^+=V\frac{c_{\lambda}}{1-\sqrt{(1-V^2)(1-c_{\lambda}^2)}}\nonumber
\\
\bar{V}^-=-V\frac{c_{\lambda}}{1+\sqrt{(1-V^2)(1-c_{\lambda}^2)}}
\end{eqnarray}
In the limit of $c_{\lambda}\rightarrow 1$ we obtain the usual
results of the special relativity with a complete symmetry
between "backward" and "forward" velocities. In general, however
these two values are not symmetrical. For example,in the limit
$V\rightarrow c_{\lambda}$ the respective values of $\bar{V}^+$
and $\bar{V}^-$ are $$\bar{V}^+ \rightarrow 1,$$ $$\bar{V}^-
\rightarrow -\frac{c_{\lambda}^2}{2-c_{\lambda}^2}$$\\

To find the law of velocity composition we notice that for
$\bar{V}$ it looks exactly like the one found in special
relativity. This allows us to write the respective law for $V$
for two particles {\bf having the same mass}. After performing
some algebra, we obtain:
\begin{equation}
\label{eq:25}
V_{1,2}=\frac{V_1+V_2-\sqrt{1-c_{\lambda}^2}(w_1+w_2)}
{1+V_1V_2+\sqrt{1-c_{\lambda}^2}[(w_1+w_2-w_1w_2)
(1+\sqrt{1-c_{\lambda}^2})-(1-c_{\lambda}^2)]}
\end{equation}
where$$w_k=1-\sqrt{1-V_k^2},$$

From (\ref{eq:25}) follows that if either $V_1$ or $V_2$ ( or
both) go to $c_{\lambda}$ the combined velocity also tends to
$c_{\lambda}$ as could be expected.\\

{\bf{II}}. Here we investigate another approach to the problem
of  the dispersion equation , based  on working directly in the
space of the boost parameter $z(\xi,\lambda)$ which depends on
both the "bare" boost parameter $\xi$ and the universal length
scale $\lambda$. We return to the Minkowski distance and write
down the respective spatio-temporal transformation from one
inertial system $K$ to a system $K'$ moving with respect to $K$
with a constant velocity $V.$

We require (in full agreement with the invariance of Minkowski
distance) this transformation to be exactly similar to the
special relativistic relations (cf.\cite{JG}):
\begin{equation}
\label{eq:A} x=x'cosh z(\xi,\lambda)+ x_0'sinh z(\xi,\lambda),~~
x_0=x'sinh z(\xi,\lambda)+x_0'cosh z(\xi,\lambda)
\end{equation}
where the boost parameter $z(\xi,\lambda)$ has to be determined.

To find parameter $z(\xi,\lambda)$ we use the fact that the
velocity of $K'$ with respect to $K$ is $$V=tanh z,$$ as follows
from (\ref{eq:A}).  Introducing this expression for $V$ into
(\ref{eq:12}) we obtain
\begin{equation}
\label{eq:B} tanh z(\xi)= \frac{sinh\xi}{cosh\xi
\sqrt{1+\lambda^2\bar{m}^2}+\bar{m}\lambda}\equiv
\frac{2e^{-m\lambda}
tanh(\frac{\xi}{2})}{1+e^{-2m\lambda}tanh^2(\frac{\xi}{2})}\equiv
tanh(2y)
\end{equation}
 where we denote $y\equiv e^{-m\lambda}
tanh(\frac{\xi}{2})$. As a result, the explicit form of the
function $z(\xi)$ is
\begin{equation}\label{eq:0}
z=2y\equiv 2tanh^{-1}[e^{-m\lambda} tanh(\frac{\xi}{2})]
\end{equation}
This function was obtained in \cite{JG} on the basis of
group-theoretical arguments, with the only difference that in
\cite{JG} was a
typographical error showing an additional factor $e^{m\lambda}$ at $tanh^{-1}$\\

Her we would like to briefly comment on the following. It has
turned out that starting with the assumption of the formal
coincidence of the spatio-temporal transformation from one
inertial system  to another with special-relativistic relations,
and using the kinematical relation ( for $V$) obtained with the
help of the momentum sector of the phase space one can arrive at
the basic relation of the algebra ( a member of a class of
algebras in ref.\cite{JL}) studied in \cite{JG}. In fact by using
Eq.{\ref{eq:13}), differentiating Eq.(\ref{eq:0}) and using the
result in transformations (\ref{eq:A}) we arrive at the basic
relation of the above algebra (in two dimensions)
$^{\cite{JG}}$:$$\delta x_0= e^{-p_0\lambda}x;~~~\delta
x=e^{-p_0\lambda}x_0.$$

The momentum $p$ is given by (\ref{eq:9}), which in terms of the
boost parameter is $$p=\bar{m}\frac{V}{\sqrt{1-V^2}}=\bar{m}cosh
z.$$ In view of this it is clear that defining energy as $p_0$
[Eq.(\ref{eq:13})] is not going to give us the relation between
$p$ and $p_0$ consistent with the $V(m=0)=1.$  The analogous case
was considered above, but there we dealt with the transformation
space defined  directly by the parameter $\xi$ and not yielding
the transformation in the form (\ref{eq:A}). The problem there
has been solved by the introduction of both $effective$
wave number and $effective$ frequency.\\

Here however we look for another transformation which will keep
the wave  number as $p$ ( in the appropriate units), but transform
$p_0$ in such a way as to produce the required $effective$
frequency. To this end we require such a effective frequency $\bar
{\omega}$ to behave like the temporal component ( vs. spatial
component $p\Leftrightarrow\bar{k}$) of the 4-vector of
coordinates defined by (\ref{eq:A})
\begin{equation}
\label{eq:C}\bar{\omega}^2=\bar{m}^2+\bar{k}^2
\end{equation}
Note that in this case ( as we have already indicated), as in the
previous case the effective mass $\bar{m},$ is the same. On the
other hand, the $effective$ wave number ($\bar{k}\Leftrightarrow
p$) and the effective frequency are different as compared to case
$I.$

Since $\bar{k}=p$ we have to find the explicit expression for the
frequency $\bar{\omega}$. Using the expression for $p$,
Eq.(\ref{eq:10}) we get
\begin{equation}
\label{eq:D}
\frac{\bar{\omega}}{\bar{m}}=\frac{cosh(m\lambda)cosh\xi+sinh(m\lambda)}
{sinh(m\lambda)cosh\xi+cosh(m\lambda)}
\end{equation}
By combining Eq.(\ref{eq:D})with the formula for $p_0$,
Eq.(\ref{eq:13}), we obtain
\begin{equation}
\label{eq:E}
{\bar{\omega}}=\frac{cosh(m\lambda)-e^{-p_0\lambda}}{\lambda}
\end{equation}

Upon substitution of $cosh(m\lambda)$ from Eq.(\ref{eq:3})
 we get the dependence of $\omega$ on $p_0$:
 \begin{equation}
 \label{eq:F} \bar\omega=
 \frac{sinh(p_0\lambda)}{\lambda}-\frac{\lambda}{2}p^2e^{p_0\lambda}
 \end{equation}

In contradistinction to case $I$, the effective frequency
$\bar{\omega}$ is finite when the momentum
$p\Leftrightarrow\bar{k}\rightarrow 1/\lambda$. This is especially
well seen if we rewrite the dispersion relation Eq.(\ref{eq:F})
 in the equivalent form:
\begin{equation}
\label{eq:G} \bar\omega=
\frac{\sqrt{cosh^2(m\lambda)-(1-\lambda^2\bar{k}^2)}}{\lambda}
\end{equation}
Let us indicate that Eq.(\ref{eq:20}) of the case $I$ has the same
form.\\

Note that if we use the Planck number , common for $k$ and
$\omega$ , we would not be able to reconcile the dispersion
relations with a massless  case where $$\bar{\omega}=\bar{k},$$
implying (according to ~ $\hbar \omega=p_0$ ~and ~ $\hbar k=p$)
that $p_0=p$ which is patently not the case. This justifies the
introduction of two Planck numbers $\hbar_{eff}^0~~and~~
\hbar_{eff}^1$ which was done earlier, Eq.(\ref{eq:19}).

We still have to choose the  dispersion equation out of
(\ref{eq:20}) and (\ref{eq:G}) on the basis of their physical
meaning. Let us consider a massless case, $m=0$. Using
Eqs.(\ref{eq:G}) and (\ref{eq:19}) we obtain that in this case
energy is $$energy = p\Leftrightarrow \bar{k}$$ and since $0\leq
p\leq 1/\lambda,$ the energy does not tend to $\infty$ with
$p\rightarrow 1/\lambda$, as predicted by relation (\ref{eq:3}).

Therefore the case $II$ is not physical in this sense, which
leaves us with case $I$ and the respective dispersion equation
(\ref{eq:20}).

\section{Uncertainty Relation}
We use this equation to explicitly obtain the value of the
$effective$ Planck number $\hbar_{eff}^1$ needed for the
comparison of the uncertainty relations here and in string theory.
Inserting the effective Planck "number" ( or rather vector),
defined according to Eq.(\ref{eq:19}), into (\ref{eq:G})  we get
\begin{eqnarray}
  \label{eq:L}
\hbar_{eff}^0=\frac{1}{p_0\lambda}\sqrt{cosh^2(m\lambda)+
(1-\lambda^2\bar{k}^2)}\nonumber \\
\hbar_{eff}^1=e^{p_0\lambda};~~\bar{k}= pe^{p_0\lambda}
\end{eqnarray}
Using this expression for $\hbar_{eff}^1$, we can compare the
uncertainty relation
\begin{equation}
\label{eq:H} \delta x\delta p \geq \hbar_{eff}^1
\end{equation}
which follow from this value and the uncertainty relation
according to the string theory. We restrict our attention to a massless case $m=0$.\\

We find that now (\ref{eq:3}) yields
\begin{equation}
\label{eq:K}
p_0=-\frac{1}{\lambda}log(1-\lambda p)
\end{equation}
Inserting (\ref{eq:L}) and (\ref{eq:K}) into (\ref{eq:H}) we
arrive at the following expression
\begin{equation}
\label{eq:27} \delta x \geq-\frac{1}{\delta p(1-\lambda \delta p)}
\end{equation}
where we use $ p \geq \delta p.$  \\

For small values of $\lambda$ we expand  (\ref{eq:27}) in Taylor
series and retain the terms up to the second power of $\lambda$.
\begin{figure}
 \begin{center}
 \includegraphics[width=6cm, height=6cm]{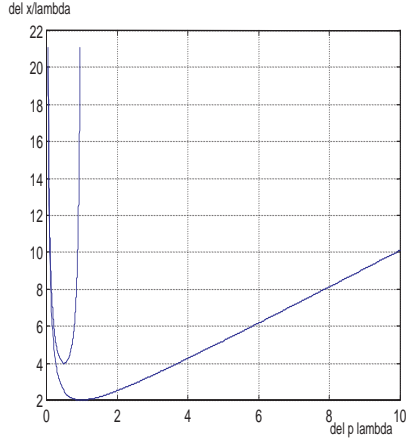}
 \caption{\small Uncertainty relations $\delta x~ vs.~ \lambda\delta p$ for a massless particle in
 string theory (lower curve) and in new relativity (upper curve);
 note the unbounded growth of $\delta x$ at the finite value of
 $\delta p =1/\lambda$}
 \end{center}
 \end{figure}
This yields the following relation:
\begin{equation}
\label{eq:28} \delta x\geq [\lambda+\frac{1}{\delta
p}+\lambda^2\delta p]
\end{equation}

The analogous relations have been obtained previously \cite{CC},
\cite{GA3}, \cite{GA4}, \cite{GA5}, although the full uncertainty
relation for a massless particle have not been investigated {\it
from a position of the dispersion relation dictating the
introduction of the effective Planck number ( or a vector) varying
with both the universal length scale and the momentum uncertainty.}\\

Comparing (\ref{eq:28}) with the string uncertainty relation
(see, for example \cite{GV}) $$\delta x\geq \frac{1}{\delta
p}+\lambda^2\delta p, $$ we notice that in the (truncated)
uncertainty relation (\ref{eq:28}) there is an extra term, linear
in $\lambda$, whereas in string theory this term is absent.
Moreover, since we consider the truncated form of (\ref{eq:27})
it does not give us the unbounded increase of $\delta x$ when
$\delta p\rightarrow \lambda$. Nonetheless, the general character
of the dependence (\ref{eq:27}) is in agreement with the string
relation, with the minimum  $\delta x_{min}= 2\lambda$, reached
at $\delta p=.5/\lambda.$ The graph with both the string
uncertainty relation and the string uncertainty relation
(\ref{eq:27})is given in Fig.1.

\section{Conclusion}
We have undertaken a task of deriving and to physically justify by
simple means the consistent ( with the velocity of  a massless
particle) two-scale relativity ( new relativity \cite{GA1})
dispersion relations.  It has turned out that in addition to the
one relation introduced in  \cite{JG} there exists another
relation which is compatible with the velocity of  a massless
particle to be $V(m=0)=1.$ However, it has turned out that this
relation is inconsistent with the the infinite energy
corresponding to the maximum value of the momentum $p=1/\lambda.$

The physical ( and even non-physical one) dispersion relation
resolves an apparent paradox arising from the fact that if one
would take the frequency to be equal to energy and wave number to
be equal to the momentum (in the appropriate units), then the
basic relation between energy and momentum cannot be translated
into a consistent dispersion relation. The seemingly puzzling
behavior of   frequency-energy and momentum-wave number has a
rather straightforward explanation.

At length scales comparable to the universal length scale
$\lambda$ the Planck constant stops to be constant and becomes a
function of both the wave number and the universal length scale.
Moreover, this function has two components (in 2 dimensions), one
corresponding to the wave number $\hbar_{eff}^1$, and another one
$\hbar_{eff}^0.$ Since we assume that the 2-scale relativity is
based upon the two universal invariants ,thus indicating that at
the scales comparable to $\lambda$ the quantities considered
constant at larger scales, become dependent on the two invariants
and dynamic variables. This restores the linear character of
frequency-energy and momentum-wave number relations.Moreover,
such an explanation is in full agreement with the string theory,
where the uncertainty relation clearly indicates a variable
character of the Planck number at scales comparable to $\lambda$.

Aknowledgement\\

I thank J.Kowalski-Glikman for his valuable comments on the draft
of this paper and to V.Granik for his help in discussing its
results.


\begin{thebibliography}{99}

\bibitem{GA1} G.Amelino-Camelia, gr-qc/0012051
\bibitem{GA2} G.Amelino-Camelia, hep-th/0012238
\bibitem{JL} J.Lukierski, H.Ruegg, and W.J.Zakrzewski, Ann.Phys.
{\bf 243}, 90 (1995)
\bibitem{LN} L.Nottale,"Fractal Space-Time and Microphysics:Towards a
Theory of Scale Relativity, World Scientific, 1993
\bibitem{CC} C.Castro, Found.Phys.Lett.,{\bf{10}}, 273
\bibitem {JG} J.Kowalski-Glikman,hep-th/0107054
\bibitem{NB} N.R.Bruno, G.Amelino-Camelia, J.Kowalski-Glikman, hep-th/0107039
\bibitem{GA3}G.Amelino-Camelia, gr-qc/9603013
\bibitem {GA4},G.Amelino-Camelia,gr-qc/9603014
\bibitem{GA5}G.Amelino-Camelia, J.Lukierski, A.Nowicki,hep-th/9706031
\bibitem{CC1} C.Castro, physics/0011040; C.Castro and A.Granik,
physics/0008222
\bibitem{GV} G.Veneziano, Europhys.Lett.{\bf 2},(1988) 199\\
D.Gross and P.Mende,Nucl.Phys. {\bf B}303(1988) 407 \\
E.Witten,Physics Today,April 1996
\end{thebibliography}
\end{document}